\documentclass{stacs_proc}
\usepackage{macros}
\usepackage{psfrag}

\begin{document}

\title[Convergence Thresholds of Newton's Method]{Convergence Thresholds of Newton's Method
 \\ for Monotone Polynomial Equations}

\author{J. Esparza}{Javier Esparza}
\author{S. Kiefer}{Stefan Kiefer}
\author{M. Luttenberger}{Michael Luttenberger}
\address{
Institut f{\"u}r Informatik \newline
Technische Universit{\"a}t M{\"u}nchen, Germany
}
\email{{esparza,kiefer,luttenbe}@in.tum.de}

\thanks{
 This work was supported by the project {\em Algorithms for Software Model Checking}
 of the {\em Deutsche Forschungsgemeinschaft (DFG)}. Part of this work was done at the Universit{\"a}t Stuttgart.}


\begin{abstract}
Monotone systems of polynomial equations (MSPEs) are systems of
fixed-point equations $X_1 = f_1(X_1, \ldots, X_n),$ $\ldots, X_n
= f_n(X_1, \ldots, X_n)$ where each $f_i$ is a polynomial with
positive real coefficients. The question of computing the least
non-negative solution of a given MSPE  $\vX = \vf(\vX)$ arises
naturally in the analysis of stochastic models such as stochastic
context-free grammars, probabilistic pushdown automata, and
back-button processes. Etessami and Yannakakis have recently
adapted Newton's iterative method to MSPEs. In a previous paper we
have proved the existence of a
threshold $k_\vf$ for strongly connected MSPEs, such that after $k_\vf$ iterations of Newton's
method each new iteration computes at least 1 new bit of the
solution. However, the proof was purely existential. In this paper
we give an upper bound for $k_\vf$ as a function of the minimal
component of the least fixed-point $\fix{\vf}$ of
$\vf(\vX)$. Using this result we show that $k_\vf$ is at most
single exponential resp.\ linear for strongly connected MSPEs
derived from probabilistic pushdown automata resp. from
back-button processes. Further, we prove the existence of a
threshold for arbitrary MSPEs after which each new iteration
computes at least $1/w2^{h}$ new bits of the solution, where $w$ and $h$
are the width and height of the DAG of strongly connected
components.
\end{abstract}

\maketitle

\stacsheading{2008}{289-300}{Bordeaux}
\firstpageno{289}

\section{Introduction}\label{sec:intro}

\noindent A {\em monotone system of polynomial equations} (MSPE for short)
has the form
$$\begin{array}{rcl}
X_1 & = & f_1(X_1, \ldots, X_n) \\
    & \vdots & \\
X_n& = & f_n(X_1, \ldots, X_n)
\end{array}$$
\noindent where $f_1, \ldots, f_n$ are polynomials with {\em
positive} real coefficients. In vector form we denote an MSPE by
$\vX = \vf(\vX)$. We call MSPEs  ``monotone'' because $\vx \leq \vx'$
implies $\vf(\vx) \leq \vf(\vx')$ for every $\vx,\vx'\in\Rp^n$. MSPEs
appear naturally in the analysis of many stochastic models, such as
context-free grammars (with numerous applications to natural language
processing \cite{ManningSchuetze:book,geman02probabilistic}, and
computational biology
\cite{Sakabikaraetal,Durbinetal:book,DowellEddy,KnudsenHein}),
probabilistic programs with procedures
\cite{EKM:prob-PDA-PCTL,BKS:pPDA-temporal,EYstacs05,
EY:RMC-LTL-complexity,EKM:prob-PDA-expectations,
EY:RMC-LTL-QUEST,EY:RMC-RMDP}, and web-surfing models with back buttons
\cite{FaginetalSTOC,Faginetal}.

By Kleene's theorem, a feasible MSPE $\vX = \vf(\vX)$ (i.e., an MSPE
with at least one solution) has a least solution $\mu\vf$; this
solution can be irrational and non-expressible by radicals.
Given an MSPE and a vector $\vv$ encoded in binary, the problem whether
$\mu\vf \leq \vv$ holds is in PSPACE and at least as hard as the
SQUARE-ROOT-SUM problem, a well-known problem of computational geometry
(see~\cite{EYstacs05,EYstacs05Extended} for more details).

For the applications mentioned above the most important
question is the efficient numerical approximation of the least solution.
Finding the least solution of a feasible system
$\vX = \vf(\vX)$
amounts to finding the least solution of $\vF(\vX) = \vzero$ for
$\vF(\vX) = \vf(\vX) - \vX$. For this we can apply
(the multivariate version of) \emph{Newton's method}
\cite{OrtegaRheinboldt:book}: starting at some $\xs{0} \in
\mathbb{R}^n$ (we use uppercase to denote variables and
lowercase to denote values), compute the sequence
$$
\xs{k+1} := \xs{k} - (\vF'(\xs{k}))^{-1}\vF(\xs{k})
$$
\noindent where $\vF'(\vX)$ is the Jacobian matrix
of partial derivatives.

While in general the method may not even be defined
($\vF'(\xs{k})$ may be singular for some~$k$), Etessami and Yannakakis proved in~\cite{EYstacs05,EYstacs05Extended}
that this is not the case for the {\em Decomposed Newton's Method (DNM)},
that decomposes the MSPE into {\em strongly connected components} (SCCs)
and applies Newton's method to them in a bottom-up fashion\footnote{A subset of variables and their associated equations form an SCC,
if the value of any variable in the subset
influences the value of all variables in the subset, 
see Section~\ref{sec:prelim} for details.}.

The results of \cite{EYstacs05,EYstacs05Extended}
provide no information on
the number of iterations needed to compute $i$ {\em valid bits}
of $\fix{\vf}$, i.e., to compute a vector $\vnu$ such that
 $\abs{\fix{\vf}_j - \nu_j} / \abs{\fix{\vf}_j} \le
2^{-i}$ for every $1 \le j \le n$. In a former paper \cite{KLE07:stoc} we
have obtained a first positive result on this problem. We have proved
that for every strongly connected MSPE $\vX = \vf(\vX)$ there exists a
threshold $k_\vf$
such that for every $i \geq 0$ the ($k_\vf+i$)-th iteration of Newton's
method has at least $i$ valid bits of $\fix{\vf}$. So, loosely speaking,
after $k_\vf$ iterations DNM is guaranteed to
compute at least 1 new bit of the solution per iteration;
we say that DNM converges {\em linearly with rate 1}.

The problem with this result is that its proof
provides no information on $k_\vf$
other than its existence. In this paper we show that
the threshold $k_\vf$ can be chosen as
$$ k_\vf = 3n^2m + 2n^2\abs{\log \mumin}$$
\noindent where $n$ is the number of equations of the MSPE, $m$ is
such that all coefficients of the MSPE
can be given as ratios of $m$-bit integers, and $\mumin$ is
the minimal component of the least solution $\mu\vf$.

It can be objected that
$k_\vf$ depends on $\fix{\vf}$, which is precisely what Newton's method
should compute. However, for MSPEs coming from stochastic models, such as
the ones listed above, we can do far better.
The following observations and results help to deal with $\mumin$:
\begin{itemize}
\item We obtain a syntactic bound on $\mumin$
for probabilistic programs with procedures (having stochastic context-free
grammars and back-button stochastic processes as special instances)
and prove that in this case $k_\vf \leq n 2^{n+2} m$.
\item We show that if every procedure has a non-zero probability of
terminating, then $k_\vf \leq 3nm$.
This condition always holds in the special case of back-button processes \cite{FaginetalSTOC,Faginetal}.
Hence, our result shows that $i$ valid bits can be computed in time \mbox{$\bigo((nm+i)\cdot n^3)$}
 in the unit cost model of Blum, Shub and Smale \cite{BlumShubSmale},
 where each single arithmetic operation over the reals can be carried out exactly and in constant time.
It was proved in~\cite{FaginetalSTOC,Faginetal} by a reduction to a semidefinite programming problem
 that $i$ valid bits can be computed in $\text{poly}(i,n,m)$-time in the classical (Turing-machine based) computation model.
We do not improve this result, because we do not have a proof that
 round-off errors (which are inevitable on Turing-machine based models) do not crucially affect the convergence of Newton's method.
But our result sheds light on the convergence of a practical method to compute $\fix{\vf}$.
\item Finally, since $\xs{k} \leq \xs{k+1} \le \fix{\vf}$ holds for every
$k \geq 0$, as Newton's method proceeds it provides better and better lower
bounds for $\mumin$ and thus for $k_\vf$.
In the paper we exhibit a MSPE for which, using this fact and
our theorem, we can prove that no component of the
solution reaches the value 1. This cannot be proved by just computing more
iterations, no matter how many.
\end{itemize}

\noindent The paper contains two further results concerning non-strongly-connected MSPEs:
Firstly, we show that DNM still converges linearly even if the MSPE has more than one SCC,
albeit the convergence rate is poorer.
Secondly, we prove that Newton's method is well-defined also for non-strongly-connected MSPEs.
Thus, it is not necessary to decompose an MSPE into its SCCs -- although
decomposing the MSPE may be preferred for efficiency reasons.
%
%
%

The paper is structured as follows.
In Section~\ref{sec:prelim} we state preliminaries and give some background
on Newton's method applied to MSPEs.
Sections~\ref{sec:conv-scMSP},~\ref{sec:conv-DNM}, and~\ref{sec:well-defined}
contain the three results of the paper.
Section~\ref{sec:pPDA} contains applications of our main result.
We conclude in Section~\ref{sec:conclusions}.
Missing proofs can be found in a technical report~\cite{EKL08:stacsTechRep}.

\section{Preliminaries}\label{sec:prelim}
\noindent In this section we introduce our notation and formalize the concepts mentioned in the introduction.

\subsection{Notation}
\noindent $\R$ and $\N$ denote the sets of real, respectively natural
numbers. We assume $0\in \N$.
$\R^n$ denotes the set of $n$-dimensional real valued
{\em column} vectors and $\Rp^n$ the subset of vectors with non-negative components.
We use bold letters for vectors, e.g. $\vx\in\R^n$, where we assume
that $\vx$ has the components $x_1,\ldots,x_n$.
Similarly, the $i^{\text{th}}$ component of a function $\vf:\R^n \to \R^n$ is denoted by~$f_i$.

$\R^{m\times n}$ denotes the set of matrices having $m$ rows and $n$ columns.
The transpose of a vector or matrix is indicated by the superscript $^\top$.
The identity matrix of $\R^{n\times n}$ is denoted by $\Id$.


The {\em formal Neumann series} of $A\in\R^{n\times n}$ is defined by $A^\ast = \sum_{k\in\N} A^k$.
It is well-known that $A^\ast$ exists if and only if the spectral radius of $A$ is less than $1$, i.e.
$\max\{ \abs{\lambda} \mid \mathbb{C}\ni\lambda \text{ is an eigenvalue of } A \} < 1$.
If $A^\ast$ exists, we have $A^\ast = ( \Id - A )^{-1}$.

The partial order $\le$ on $\R^n$ is defined as usual
  by setting $\vx \le \vy$ if $x_i \le y_i$ for all $1 \le i \le n$.
By $\vx < \vy$ we mean $\vx \le \vy$ and $\vx \neq \vy$.
Finally, we write $\vx \prec \vy$ if $x_i < y_i$ in every component.

We use $X_1,\ldots,X_n$ as variable identifiers and arrange them into the vector $\vX$.
In the following $n$ always denotes the number of variables, i.e. the dimension of $\vX$.
While $\vx,\vy,\ldots$ denote arbitrary elements in $\R^n$, resp. $\Rp^n$,
 we write $\vX$ if we want to emphasize that a function is given w.r.t.\ these variables.
Hence, $\vf(\vX)$ represents the function itself, whereas $\vf(\vx)$ denotes its value for $\vx\in\R^n$.

If $Y$ is a set of variables and $\vx$ a vector,
 then by $\vx_Y$ we mean the vector obtained by restricting $\vx$ to the components in $Y$.

The {\em Jacobian} of a differentiable function $\vf(\vX)$ with $\vf : \R^n \to \R^m$ is the matrix $\vf'(\vX)$ given by
\[
  \vf'(\vX) =
  \begin{pmatrix}
  \pd{f_1}{X_1} & \ldots & \pd{f_1}{X_n} \\
     \vdots     &        & \vdots\\
  \pd{f_m}{X_1} & \ldots & \pd{f_m}{X_n} \\
  \end{pmatrix} \:.
\]
\subsection{Monotone Systems of Polynomials}
\begin{definition}
A function $\vf(\vX)$ with $\vf : \Rp^n \to \Rp^n$ is a
{\em monotone system of polynomials (MSP)},
  if every component $f_i(\vX)$ is a polynomial in the variables $X_1,\ldots,X_n$ with coefficients in $\Rp$.
We call an MSP $\vf(\vX)$ {\em feasible} if $\vy = \vf(\vy)$
for some $\vy \in \Rp^n$.
\end{definition}
\begin{fact}
Every MSP $\vf$ is monotone on $\Rp^n$, i.e. for $\vzero \le \vx \le \vy$ we have $\vf(\vx) \le \vf(\vy)$.
\end{fact}
\noindent Since every MSP is continuous, Kleene's fixed-point theorem
(see e.g. \cite{Kui}) applies.
\begin{theorem}[Kleene's fixed-point theorem]\label{thm:kleene}
Every feasible MSP $\vf(\vX)$ has a least fixed point $\mu\vf$ in $\Rp^n$
i.e., $\mu\vf = \vf(\mu\vf)$ and, in addition, $\vy = \vf(\vy)$ implies
$\mu\vf \leq \vy$.
Moreover, the sequence $(\ks{k}_{\vf})_{k\in\N}$ with $\ks{0}_{\vf}:= \vzero$, and $\ks{k+1}_{\vf} := \vf( \ks{k}_{\vf} ) = \vf^{k+1}(\vzero)$ is monotonically increasing with respect
to $\leq$ (i.e. $\ks{k}_{\vf} \leq \ks{k+1}_{\vf})$
and converges to~$\mu\vf$.
\end{theorem}
\noindent In the following we call $(\ks{k}_{\vf})_{k\in\N}$ the {\em Kleene sequence}
of $\vf(\vX)$, and drop the subscript whenever $\vf$ is clear from the context.
Similarly, we sometimes write $\vmu$ instead of $\fix{\vf}$.

A variable $X_i$ of an MSP $\vf(\vX)$ is {\em productive} if $\ksc{k}_i > 0$
for some  $k\in\N$. An MSP is {\em clean} if all its variables are productive.
It is easy to see that $\ksc{n}_i = 0$ implies $\ksc{k}_i = 0$ for all
$k\in\N$. As for context-free grammars we can determine
all productive variables in time linear in the size of~$\vf$.

\begin{notation}
 In the following, we always assume that an MSP $\vf$ is clean and feasible.
 I.e., whenever we write ``MSP'', we mean ``clean and feasible MSP'',
 unless explicitly stated otherwise.
\end{notation}
%

\noindent For the formal definition of the
{\em Decomposed Newton's Method (DNM)} (see also Section~\ref{sec:intro})
we need the notion of {\em dependence} between variables.
\begin{definition}
Let $\vf(\vX)$ be an MSP. $X_i$ {\em depends directly} on $X_k$,
denoted by $X_i \trianglelefteq X_k$, if $\pd{f_i}{X_k}(\vX)$ is not the
zero-polynomial. $X_i$ {\em depends} on $X_k$ if $X_i \trianglelefteq^{\ast} X_k$, where $\trianglelefteq^\ast$ is the reflexive transitive closure of $\trianglelefteq$. An MSP is {\em strongly connected} (short: an {\em scMSP})
if all its variables depend on each other.
\end{definition}
\noindent Any MSP can be decomposed into strongly connected components (SCCs),
 where an SCC $S$ is a maximal set of variables such that each variable in $S$ depends on each other variable in $S$.
The following result for strongly connected MSPs was proved in \cite{EYstacs05,EYstacs05Extended}:
\begin{theorem}
\label{scMSP:th}
Let $\vf(\vX)$ be an scMSP and define the Newton operator
$\Ne_{\vf}$ as follows
$$
\Ne_{\vf}(\vX) = \vX + (\Id-\vf'(\vX))^{-1} ( \vf(\vX) - \vX ) \; .
$$
\noindent We have:
  (1) $\Ne_{\vf}(\vx)$ is defined for all $\vzero \le \vx \prec \mu\vf$ (i.e., $(\Id - \vf'(\vx))^{-1}$ exists).
  Moreover, $\vf'(\vx)^\ast = \sum_{k\in\N} \vf'(\vx)^k$ exists for all $\vzero \le\vx\prec \mu\vf$,
   and so $\Ne_{\vf}(\vX) = \vX + \vf'(\vX)^\ast(\vf(\vX)-\vX)$.
  (2) The Newton sequence $(\ns{k}_{\vf})_{k\in\N}$ with $\ns{k} = \Ne_{\vf}^k(\vzero)$ is monotonically increasing,
    bounded from above by $\mu\vf$ (i.e. $\ns{k} \le \vf(\ns{k}) \le \ns{k+1} \prec \mu\vf$), and converges to $\mu\vf$.
\end{theorem}
%

\noindent DNM works by substituting the variables of lower SCCs by corresponding
Newton approximations that were obtained earlier.
%
%

\section{A Threshold for scMSPs}\label{sec:conv-scMSP}

\noindent In this section we obtain a threshold after which
DNM is guaranteed to converge linearly with rate 1.

We showed in \cite{KLE07:stoc} that for worst-case results on the
convergence of Newton's method it is enough to consider
\emph{quadratic} MSPs, i.e., MSPs whose monomials have degree at most 2.
The reason is that any MSP (resp.\ scMSP) $\vf$ can be transformed
into a quadratic MSP (resp.\ scMSP) $\widetilde \vf$
by introducing auxiliary variables.
This transformation is very similar to the transformation of a
context-free grammar into Chomsky normal form. The transformation does not accelerate
DNM, i.e., DNM on $\vf$ is at least as fast
(in a formal sense) as DNM on $\widetilde \vf$, and so
for a worst-case analysis, it suffices to
consider quadratic systems. We refer the reader to \cite{KLE07:stoc}
for details.

We start by defining the notion of ``valid bits''.

\begin{definition}\label{def:i-valid-bits}
Let $\vf(\vX)$ be an MSP. 
A vector $\vnu$ has {\em $i$ valid bits} of the least fixed point
$\fix{\vf}$ if $\abs{\fix{\vf}_j - \nu_j} / \abs{\fix{\vf}_j} \le
2^{-i}$ for every $1 \le j \le n$.
\end{definition}

\noindent In the rest of the section we prove the following:

\begin{theorem}\label{thm:estimate}
 Let $\vf(\vX)$ be a quadratic scMSP.
 Let $\cmin$ be the smallest nonzero coefficient of $\vf$ and let $\mumin$
and $\mumax$ be the minimal and maximal component of $\fix{\vf}$,
respectively. Let $$ \displaystyle{k_\vf = n \cdot \log \left( \frac{\mumax}{\cmin \cdot \mumin \cdot \min\{\mumin, 1\}}\right)}  \:.$$
 Then $\ns{\lceil k_\vf \rceil + i}$ has $i$ valid bits of $\fix{\vf}$
for every $i \geq 0$.
\end{theorem}

\noindent Loosely speaking, the theorem states that after $k_\vf$ iterations
of Newton's method, every subsequent iteration guarantees at least
one more valid bit. It may be objected that $k_\vf$ depends on the
least fixed point $\mu\vf$, which is precisely what Newton's
method should compute. However, in the next section we show that
there are important classes of MSPs (in fact, those which
motivated our investigation), for which bounds on
$\mumin$ can be easily obtained.

The following corollary is weaker than Theorem \ref{thm:estimate}, but less technical in that it avoids a dependence on $\mumax$ and $\cmin$.

\begin{corollary}\label{cor:estimate-size}
Let $\vf(\vX)$ be a quadratic scMSP of dimension $n$ whose coefficients are given as ratios of $m$-bit integers.
Let $\mumin$ be the minimal component of $\mu\vf$.
Let
 $ k_\vf = 3n^2m + 2n^2\abs{\log \mumin} \:.$
 Then $\ns{\lceil k_\vf \rceil + i}$ has at least $i$ valid
bits of $\fix{\vf}$ for every $i \geq 0$.
\end{corollary}

\noindent Corollary~\ref{cor:estimate-size} follows from Theorem~\ref{thm:estimate}
 by a suitable bound on $\mumax$ in terms of $\cmin$ and $\mumin$ \cite{EKL08:stacsTechRep}
(notice that, since $\cmin$ is the quotient of two $m$-bit integers, we have $\cmin \ge 1/2^m$).

In the rest of the section we sketch the proof of Theorem~\ref{thm:estimate}.
The proof makes crucial use of
vectors $\vd \succ \vzero$ such that $\vd \ge \vf'(\mu\vf)
\vd$. We call a vector satisfying these two conditions a {\em cone
vector of $\vf$} or, when $\vf$ is clear from the context, just a
cone vector.

In a previous paper we have shown that if the matrix \ $(\Id - \vf'(\mu\vf))$\ is singular,
then $\vf$ has a cone vector (\cite{KLE07:stoc}, Lemmata 4 and 8).
As a first step towards the proof of Theorem~\ref{thm:estimate} we show the
following stronger proposition.
\begin{proposition}\label{prop:ex-d}
 Any scMSP has a cone vector.
\end{proposition}

\noindent
To a cone vector $\vd = (d_1, \ldots, d_n)$ we
associate two parameters, namely the maximum and the minimum of
the ratios $\fix{\vf}_1/d_1, \fix{\vf}_2/d_2, \ldots,
\fix{\vf}_n/d_n$, which we denote by $\lmax$ and $\lmin$,
respectively. The second step consists of showing (Proposition \ref{prop:blubblubblub}) that given a cone vector $\vd$,
the threshold $k_{\vf,\vd} = \log (\lmax / \lmin)$ satisfies the
same property as $k_\vf$ in Theorem~\ref{thm:estimate},
i.e., $\ns{\lceil k_{\vf,\vd} \rceil + i}$ has $i$ valid bits of $\fix{\vf}$ for every $i \geq 0$. This follows rather easily from the following fundamental
property of cone vectors: a cone vector leads to
an upper bound on the error
of Newton's method.
%
%
%
\begin{lemma}\label{lem:blubblub}
Let $\vd$ be a cone vector of an MSP $\vf$ and let $\lmax = \max\{
\frac{\mu\vf_i}{d_i} \}$. Then
$$
 \mu\vf - \ns{k} \le 2^{-k} \lmax\ \vd.
$$
\end{lemma}
\noindent {\em Proof Idea.}
Consider the ray $\vg(t)=\mu\vf - t \vd$
starting in $\fix{\vf}$ and headed in the direction $-\vd$ (the dashed line in the picture below). It is easy to see that
$\vg(\lmax)$ is the intersection of $\vg$ with an axis which is located farthest from $\mu\vf$. One can then prove $\vg(\frac{1}{2}\lmax) \leq \ns{1}$,
where $\vg(\frac{1}{2}\lmax)$ is the point of the ray equidistant from
$\vg(\lmax)$ and $\mu\vf$. By repeated application of this argument
one obtains $\vg(2^{-k} \lmax) \le \ns{k}$ for all $k\in\N$.

The following picture shows the Newton iterates $\ns{k}$ for $0 \le k \le 2$ (shape: $\boldsymbol{\times}$) and the corresponding
points $\vg(2^{-k} \lmax)$ (shape: $\boldsymbol{+}$) located on the ray $\vg$.
Notice that $\ns{k} \ge \vg(2^{-k}\lmax)$. \qed

\begin{center}
{
 \psfrag{X1 = f1(X1,X2)}{$X_1 = f_1(\vX)$}
 \psfrag{X2 = f2(X1,X2)}{$X_2 = f_2(\vX)$}
 \psfrag{muf}{$\fix{\vf}=\vg(0)$}
 \psfrag{0}{$0$}
 \psfrag{0.2}{}
 \psfrag{0.4}{}
 \psfrag{0.6}{}
 \psfrag{\2610.4}{$-0.4$}
 \psfrag{\2610.2}{$-0.2$}
 \psfrag{02h}{$0.2$}
 \psfrag{04h}{$0.4$}
 \psfrag{06h}{$0.6$}
 \psfrag{02v}{$0.2$}
 \psfrag{X1}{$X_1$}
 \psfrag{X2}{$X_2$}
 \psfrag{glmax}{$\vg(\lmax)$}
  \scalebox{0.8}{ \includegraphics{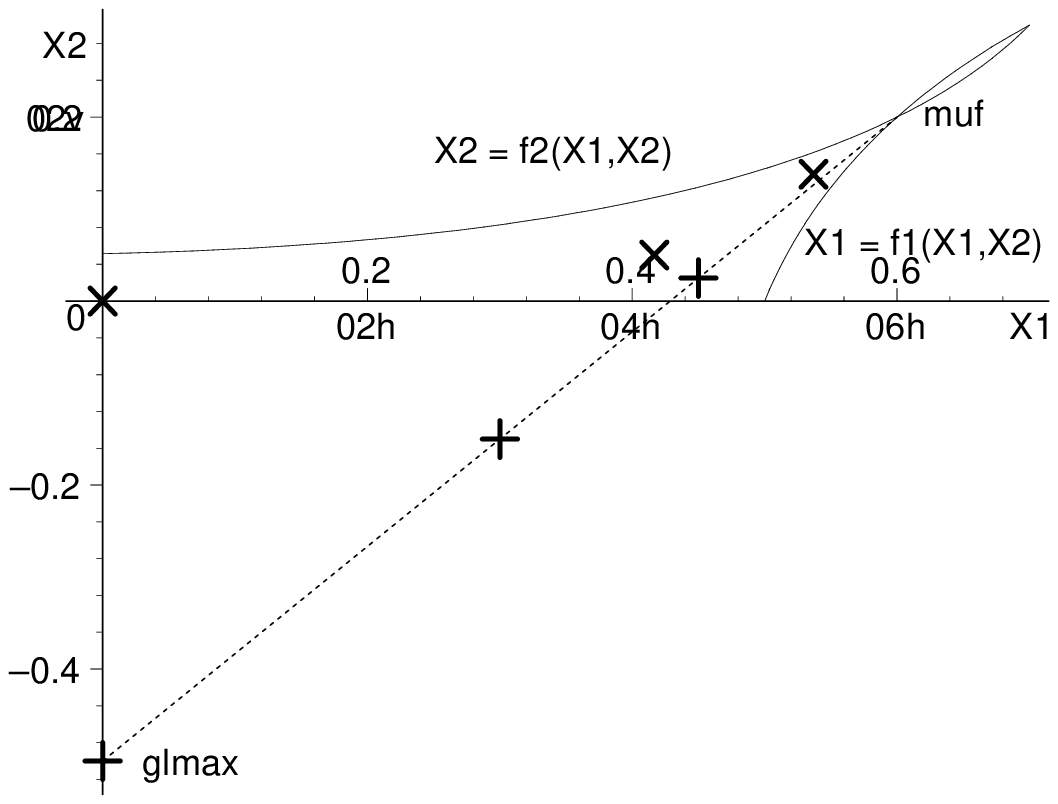}}
}
\end{center}

\noindent Now we easily obtain:
\begin{proposition}
\label{prop:blubblubblub} Let $\vf(\vX)$ be an scMSP and let $\vd$
be a cone vector of $\vf$. Let $k_{\vf,\vd} = \log
\frac{\lmax}{\lmin}$, where $\lmax = \max_j
\frac{\fix{\vf}_j}{d_j}$ and $\lmin = \min_j
\frac{\fix{\vf}_j}{d_j}$. Then $\ns{\lceil k_{\vf,\vd} \rceil +
i}$ has at least $i$ valid bits of $\fix{\vf}$ for every $i \geq
0$.
\end{proposition}

\noindent We now proceed to the third and final step.
We have the problem that $k_{\vf,\vd}$ depends on the cone vector $\vd$,
 about which we only know that it exists (Proposition~\ref{prop:ex-d}).
We now sketch how to obtain the threshold $k_\vf$ claimed in Theorem~\ref{thm:estimate}, which is independent of any
cone vectors.

Consider Proposition~\ref{prop:blubblubblub} and
 let $\lmax = \frac{\fix{\vf}_i}{d_i}$ and $\lmin = \frac{\fix{\vf}_j}{d_j}$.
We have $k_{\vf,\vd} = \log \left(\frac{d_j}{d_i} \cdot \frac{\fix{\vf}_i}{\fix{\vf}_j}\right)$.
The idea is to bound $k_{\vf,\vd}$ in terms of $\cmin$.
We show that if $k_{\vf,\vd}$ is very large, then there
 must be variables $X,Y$ such that $X$ depends on $Y$ only via a monomial
 that has a very small coefficient, which implies that $\cmin$ is very small.


 \section{Stochastic Models}\label{sec:pPDA}

 \noindent As mentioned in the introduction, several problems
 concerning stochastic models can be reduced to problems
 about the least solution $\fix{\vf}$ of an MSPE $\vf$. In these cases,
 $\fix{\vf}$ is a vector of probabilities, and so $\mumax \leq 1$.
 Moreover, we can obtain information on $\mumin$, which
 leads to bounds on the threshold $k_\vf$.

 \subsection{Probabilistic Pushdown Automata}
 \noindent Our study of MSPs was initially motivated by the verification of
 probabilistic
 pushdown automata. A \emph{probabilistic pushdown automaton (pPDA)}
 is a tuple
 $\pPDA = (Q,\Gamma,\delta,\Prob)$
 where $Q$ is a finite set of \emph{control states}, $\Gamma$ is a
 finite \emph{stack alphabet},
 $\delta \subseteq Q \times \Gamma \times Q \times \Gamma^*$ is
 a  finite \emph{transition relation} (we write $pX \tran{} q \alpha$ instead
 of $(p,X,q,\alpha) \in \delta$), and $\Prob$ is a function which to
 each transition $pX \tran{} q\alpha$ assigns
 its probability $\Prob(pX \tran{} q\alpha) \in (0,1]$ so that for
 all $p \in Q$ and $X \in \Gamma$ we have
 $\sum_{pX \tran{} q\alpha} \Prob(pX \tran{} q\alpha) = 1$.
 We write $pX \tran{x} q\alpha$ instead of $\Prob(pX \tran{} q\alpha) = x$.
 A {\em configuration} of $\pPDA$ is a pair $qw$, where $q$ is a control
 state and $w \in \Gamma^*$ is a {\em stack content}.
 A probabilistic pushdown automaton $\pPDA$
 naturally induces a possibly infinite Markov chain with the
 configurations as states and transitions given by:
 $pX \beta \tran{x} q \alpha \beta$ for every $\beta \in \Gamma^*$
 if{}f $pX \tran{x} q\alpha$. We assume w.l.o.g. that if
 $pX \tran{x} q\alpha$ is a transition then $|\alpha|\leq 2$.

 \newcommand{\vmax}{\mumax}
 \newcommand{\vmin}{\mumin}
 pPDAs and the equivalent model of recursive Markov chains have been
 very thoroughly studied
 \cite{EKM:prob-PDA-PCTL,BKS:pPDA-temporal,EYstacs05,EY:RMC-LTL-complexity,EKM:prob-PDA-expectations,EY:RMC-LTL-QUEST,EY:RMC-RMDP}.
 These papers have shown that the key to the
 analysis of pPDAs are the {\em termination probabilities} $\Pro{pXq}$,
 where $p$ and $q$ are states, and $X$ is a stack letter, defined as follows
 (see e.g. \cite{EKM:prob-PDA-PCTL} for a more formal definition): $\Pro{pXq}$ is the
 probability that, starting at the configuration $pX$, the pPDA
 eventually reaches the configuration $q\varepsilon$ (empty stack). It is not
 difficult to show that the vector of termination probabilities is the least fixed point of the MSPE containing the equation
 $$
       \Pro{pXq}  =  \sum_{pX \tran{x} rYZ}
           x \cdot \sum_{t \in Q} \Pro{rYt} \cdot \Pro{tZq}
         \quad + \quad \sum_{pX \tran{x} rY}
           x \cdot \Pro{rYq}
         \quad + \quad  \sum_{pX \tran{x} q \varepsilon} x
 $$

 \noindent for each triple $(p,X,q)$. Call this quadratic MSPE the {\em termination
 MSPE} of the pPDA (we assume that termination MSPEs are clean,
 and it is easy to see that they are always feasible). We
 immediately have that if $\vX = \vf(\vX)$ is a termination
 MSP, then $\vmax \leq 1$. We also obtain a lower bound on
 $\vmin$:

 \begin{lemma}\label{lem:lower-bound-mu}
 Let $\vX = \vf(\vX)$ be a termination MSPE with $n$ variables.
 Then $\vmin \geq \cmin^{(2^{n+1}-1)}$.
 \end{lemma}

 \noindent Together with Theorem~\ref{thm:estimate} we get the
 following exponential bound for $k_\vf$.

 \begin{proposition}\label{prop:lower-bound-exp}
 Let $\vf$ be a strongly connected termination MSP with $n$ variables
 and whose coefficients are expressed as ratios of $m$-bit numbers. Then
 $k_\vf \leq n 2^{n+2} m$.
 \end{proposition}


 \noindent We conjecture that there is a lower bound on $k_\vf$ which is exponential in $n$ for the following reason.
 We know a family $(\vf^{(n)})_{n=1,3,5,\ldots}$ of strongly connected MSPs
  with $n$ variables and irrational coefficients
  such that $\cmin^{(n)} = \frac{1}{4}$ for all $n$ and $\mumin^{(n)}$ is double-exponentially small in $n$.
 Experiments suggest that $\Theta(2^n)$ iterations are needed for the first bit of $\fix{\vf^{(n)}}$,
  but we do not have a proof.


 \subsection{Strict pPDAs and Back-Button Processes}

 \noindent
 A pPDA is {\em strict} if for all $pX \in Q  \times \Gamma$ and all $q \in Q$ the
 transition relation contains a pop-rule $pX \tran{x} q \epsilon$
 for some $x > 0$.
 %
 Essentially, strict pPDAs model
 programs in which every procedure has at least one terminating execution that
 does not call any other procedure.
 The termination MSP of a strict pPDA is of the form $\vb(\vX,\vX)+\vl\vX +\vc$
 for $\vc \succ \vzero$. So we have $\fix{\vf} \ge \vc$,
 which implies $\vmin \geq \cmin$. Together with Theorem~\ref{thm:estimate} we get:

 \begin{proposition}\label{prop:lower-bound-backbutton}
 Let $\vf$ be a strongly connected termination MSP with $n$ variables
 and whose coefficients are expressed as ratios of $m$-bit numbers.
 If $\vf$ is derived from a strict pPDA, then
 $k_\vf \leq 3nm$.
 \end{proposition}

 \noindent Since in most applications $m$ is small, we obtain an excellent convergence
 threshold.

 In \cite{FaginetalSTOC,Faginetal} Fagin et al.\ introduce
 a special class of strict pPDAs called {\em back-button processes}:
 in a back-button process there is only one control state $p$ 
 , and any rule is of the form $pA \tran{b_A} p\varepsilon$ or $pA \tran{l_{AB}} pBA$.
 So the stack corresponds to a path through a finite graph with $\Gamma$ as set of nodes and edges
 $A \to B$ for $pA \tran{l_{AB}} pBA$.

 In \cite{FaginetalSTOC,Faginetal} back-button processes are used to model the behaviour
 of web-surfers: $\Gamma$ is the set of web-pages,
 $l_{AB}$ is the probability that a web-surfer uses a link from page $A$ to page $B$,
 and $b_{A}$ is the probability that the surfer pushes the ``back''-button of the web-browser
 while visiting $A$.
 %
 %
 Thus, the termination probability $[pAp]$
 is simply the probability that, if $A$ is on top of the stack, $A$ is eventually popped from
 the stack. The termination probabilities are the least solution of the MSPE consisting of the equations
 \[
   \Pro{pAp} \quad = \quad b_A + \displaystyle{\sum_{pA \tran{l_{AB}} pBA} l_{AB}\Pro{pBp}\Pro{pAp}}
       \quad = \quad b_A + \Pro{pAp} \displaystyle{\sum_{pA \tran{l_{AB}} pBA}} l_{AB} \Pro{pBp}.
 \]
 %
 %
 %
 %

 \subsection{An Example}
 \noindent As an example of application of Theorem~\ref{thm:estimate} consider
 the following scMSPE $\vX = \vf(\vX)$.
 \[
  \begin{pmatrix}
   X_1 \\ X_2 \\ X_3
  \end{pmatrix}
  =
  \begin{pmatrix}
   0.4 X_2 X_1 + 0.6 \\
   0.3 X_1 X_2 + 0.4 X_3 X_2 + 0.3 \\
   0.3 X_1 X_3 + 0.7
  \end{pmatrix}
 \]
 \noindent The least solution of the system gives the revocation probabilities
 of a back-button process with three web-pages. For instance, if
 the surfer is at page 2 it can choose between following links
 to pages 1 and 3 with probabilities 0.3 and 0.4, respectively, or pressing the
 back button with probability 0.3.

 We wish to know if any of the revocation probabilities is equal to 1. Performing
 $14$ Newton steps (e.g.\ with Maple) yields an approximation $\ns{14}$ to the
 termination probabilities with
 \[
  \begin{pmatrix}
   0.98 \\
   0.97 \\
   0.992
  \end{pmatrix}
  \le
  \ns{14}
  \le
  \begin{pmatrix}
   0.99 \\
   0.98 \\
   0.993
  \end{pmatrix} \:.
 \]
 We have $\cmin = 0.3$. In addition, since Newton's method
 converges to $\fix{\vf}$ from below, we know $\mumin \ge 0.97$.
 Moreover, $\mumax \le 1$, as $\vone = \vf(\vone)$
 and so $\fix{\vf} \leq \vone$. Hence $k_\vf \le 3 \cdot
 \log \frac{1}{0.97 \cdot 0.3 \cdot 0.97} \le 6$.
 Theorem~\ref{thm:estimate} then implies that $\ns{14}$ has (at
 least) 8 valid bits of $\fix{\vf}$. As $\fix{\vf} \leq \vone$, the
 absolute errors are bounded by the relative errors, and since
 $2^{-8} \le 0.004$ we know:
 \[
  \fix{\vf} \prec
  \ns{14}
  +
  \begin{pmatrix}
   2^{-8}\\
   2^{-8}\\
   2^{-8}
  \end{pmatrix}
  \prec
  \begin{pmatrix}
   0.994\\
   0.984\\
   0.997\\
  \end{pmatrix}
  \prec
  \begin{pmatrix}
   1\\
   1\\
   1\\
  \end{pmatrix}
 \]
 So Theorem~\ref{thm:estimate} gives a proof that all 3 revocation
 probabilities are strictly smaller than~$1$.

 %

 \section{Linear Convergence of the Decomposed Newton's Method}\label{sec:conv-DNM}
 \noindent Given a strongly connected MSP $\vf$,
 Theorem~\ref{thm:estimate} states that, if we have computed
 $k_\vf$ preparatory iterations of Newton's method, then
 after $i$ additional iterations we can be sure to have computed
 at least $i$ bits of $\fix{\vf}$.
 We call this linear convergence with rate $1$.
 %
 %
 Now we show that DNM, which handles non-strongly-connected MSPs, converges linearly as well.
 We also give an explicit convergence rate.

 Let $\vf(\vX)$ be any quadratic MSP (again we assume {\em quadratic} MSPs throughout this section),
 and let $h(\vf)$ denote the height of the DAG of strongly connected components (SCCs).
 The convergence rate of DNM crucially depends on this height:
 In the worst case one needs asymptotically $\Theta(2^{h(\vf)})$ iterations in each component per bit,
  assuming one performs the same number of iterations in each component.

 To get a sharper result,
  we suggest to perform a different number of iterations in each SCC, depending on its {\em depth}.
 The depth of an SCC $S$ is the length of the longest path in the DAG of SCCs from $S$ to a top SCC.

 In addition, we use the following notation.
 \newcommand{\comp}{{\it comp}}
 For a depth $t$, we denote by $\comp(t)$ the set of SCCs of depth $t$.
 Furthermore we define $C(t)$ := $\bigcup \comp(t)$ and $C_{>}(t)$ := $\bigcup_{t' > t} C(t')$ and, analogously, $C_{<}(t)$.
 We will sometimes write $\vv_t$ for $\vv_{C(t)}$ and $\vv_{>t}$ for $\vv_{C_{>}(t)}$ and $\vv_{<t}$ for $\vv_{C_{<}(t)}$,
  where $\vv$ is any vector.

 Figure~\ref{fig:decompNewton} shows the Decomposed Newton's Method (DNM) for computing an approximation $\vnu$ for $\fix{\vf}$,
  where $\vf(\vX)$ is any quadratic MSP.
 The authors of \cite{EYstacs05} recommend to run Newton's Method in each SCC $S$ until
 ``approximate solutions for $S$ are considered `good enough' ''.
 Here we suggest to run Newton's Method in each SCC $S$ for a number of steps that depends (exponentially) on the depth of $S$
  and (linearly) on a parameter $j$ that controls the number of iterations (see Figure~\ref{fig:decompNewton}).

 \begin{figure} [ht]
     \centering
     \fbox{\parbox{9.5cm}{ \flushleft 
         \textbf{function} DNM $\left(\vf, j\right)$\\
         /* \emph{The parameter $j$ controls the number of iterations.} */\\
         \textbf{for} $t$ \textbf{from} $h(\vf)$ \textbf{downto} $0$ \\
         \ind \textbf{forall} $S \in \comp(t)$ \hspace{1mm} /* \emph{all SCCs $S$ of depth $t$} */ \\
         \indd $\vnu_S$ := $\Ne_{\vf_S}^{j\cdot 2^t}(\vzero)$ \hspace{3mm} /* \emph{$j \cdot 2^t$ iterations} */ \\
         \indd /* \emph{apply $\vnu_S$ in the depending SCCs } */ \\
         \indd $\vf_{<t}(\vX)$ := $\vf_{<t}(\vX) [\vX_S/\vnu_S]$ \\
         \textbf{return} $\vnu$  \\ 
     }}
     \caption{Decomposed Newton's Method (DNM) for computing an approximation $\vnu$ of $\fix{\vf}$ } 
     \label{fig:decompNewton}
 \end{figure}

 \noindent Recall that $h(\vf)$ was defined as the height of the DAG of SCCs.
 Similarly we define the width $w(\vf)$ to be $\max_t \abs{\comp(t)}$.
 Notice that $\vf$ has at most $(h(\vf) + 1) \cdot w(\vf)$ SCCs.
 We have the following bound on the number of iterations run by DNM.
 \begin{proposition}\label{prop:DNM-how-many-iterations}
  The function \textup{DNM}$\left(\vf, j\right)$ of Fig.~\ref{fig:decompNewton} runs at most
   $j \cdot w(\vf) \cdot 2^{h(\vf) + 1}$ iterations of Newton's method.
 \end{proposition}

 \noindent We will now analyze the convergence behavior of DNM 
  asymptotically (for large $j$).
 Let $\Ds{j}_S$ denote the error in $S$ when running DNM with parameter $j$,
  i.e., $\Ds{j}_S$ := $\vmu_S - \ns{j}_S$.
 Observe that the error $\Ds{j}_t$ can be understood as the sum of two errors:
 \[ \Ds{j}_t = \vmu_t - \ns{j}_t = (\vmu_t - \widetilde{\vmu_t}^{(j)}) + (\widetilde{\vmu_t}^{(j)} - \ns{j}_t) \:, \]
 where $\widetilde{\vmu_t}^{(j)}$ := $\fix{\big(\vf_t(\vX)[\vX_{>t} / \ns{j}_{>t}]\big)}$,
  i.e., $\widetilde{\vmu_t}^{(j)}$ is the least fixed point of $\vf_t$ after the approximations from the lower SCCs have been applied.
 So, $\Ds{j}_t$ consists of the {\em propagation error} $(\vmu_t - \widetilde{\vmu_t}^{(j)})$
   and the newly inflicted {\em approximation error} $(\widetilde{\vmu_t}^{(j)} - \ns{j}_t)$.

 The following lemma, technically non-trivial to prove, gives a bound on the propagation error.
 \begin{lemma}[Propagation error]\label{lem:propagation-error}
  There is a constant $c > 0$ such that
  \[ \norm{\vmu_t - \widetilde{\vmu_t}} \le c \cdot \sqrt{\norm{\vmu_{>t} - \vnu_{>t}}}
  \]
  holds for all $\vnu_{>t}$ with $\vzero \le \vnu_{>t} \le \vmu_{>t}$,
   where $\widetilde{\vmu_t} = \fix{\big(\vf_t(\vX)[\vX_{>t} / \vnu_{>t}]\big)}$.
 \end{lemma}
 \noindent Intuitively, Lemma~\ref{lem:propagation-error} states that if $\vnu_{>t}$ has $k$ valid bits of $\vmu_{>t}$,
  then $\widetilde{\vmu_t}$ has roughly $k/2$ valid bits of $\vmu_t$.
 In other words, (at most) one half of the valid bits are lost on each level of the DAG due to the propagation error.

 The following theorem assures that after combining the propagation error and the approximation error,
  DNM still converges linearly.
 \begin{theorem}\label{thm:dnm-error}
  Let $\vf$ be a quadratic MSP.
  Let $\ns{j}$ denote the result of calling \textup{DNM}$(\vf, j)$ (see Figure~\ref{fig:decompNewton}).
  Then there is a $k_\vf \in \Nat$ such that $\ns{k_\vf + i}$ has at least $i$ valid bits of $\fix{\vf}$ for every $i \ge 0$.
 \end{theorem}

 \noindent We conclude that increasing $i$ by one gives us asymptotically at least one additional bit in each component
  and, by Proposition~\ref{prop:DNM-how-many-iterations}, costs $w(\vf) \cdot 2^{h(\vf) + 1}$ additional Newton iterations.

 In the technical report~\cite{EKL08:stacsTechRep} we give an example that shows that
 the bound above is essentially optimal in the sense that an exponential (in $h(\vf)$) number of iterations
 is in general needed to obtain an additional bit.
 %
 %

 \section{Newton's Method for General MSPs}\label{sec:well-defined}

 \noindent Etessami and Yannakakis \cite{EYstacs05} introduced DNM
  because they could show that the matrix inverses used by Newton's method exist
  if Newton's method is run on each SCC separately (see Theorem~\ref{scMSP:th}).

 It may be surprising that the matrix inverses used by Newton's method
  exist even if the MSP is {\em not} decomposed.
 More precisely one can show the following theorem, see~\cite{EKL08:stacsTechRep}.

 \begin{theorem} \label{thm:well-defined}
 Let $\vf(\vX)$ be any MSP, not necessarily strongly connected.
 Let the Newton operator $\Ne_{\vf}$ be defined as before:
 \[
 \Ne_{\vf}(\vX) = \vX + (\Id-\vf'(\vX))^{-1} ( \vf(\vX) - \vX )
 \]
 \noindent Then the Newton sequence
 $(\ns{k}_{\vf})_{k\in\N}$ with $\ns{k} = \Ne_{\vf}^k(\vzero)$ is
 well-defined (i.e., the matrix inverses exist),
 monotonically increasing, bounded from above by $\mu\vf$ (i.e.
 $\ns{k} \le \ns{k+1} \prec \mu\vf$), and converges to $\mu\vf$.
 \end{theorem}

 \noindent By exploiting Theorem~\ref{thm:dnm-error} and Theorem~\ref{thm:well-defined} one can show the following theorem
  which addresses the convergence {\em speed} of Newton's Method in general.

 \begin{theorem}\label{thm:conv-speed-general}
  Let $\vf$ be any quadratic MSP.
  Then the Newton sequence $(\ns{k})_{k \in \Nat}$ is well-defined and converges linearly to $\fix{\vf}$.
  More precisely, there is a $k_\vf \in \Nat$ such that $\ns{k_\vf + i \cdot (h(\vf) + 1) \cdot 2^{h(\vf)}}$
   has at least $i$ valid bits of $\fix{\vf}$ for every $i \ge 0$.
 \end{theorem}
 \noindent Again, the $2^{h(\vf)}$ factor cannot be avoided in general
 as shown by an example in \cite{EKL08:stacsTechRep}.

 \section{Conclusions}\label{sec:conclusions}
 \noindent We have proved a threshold $k_\vf$ for strongly connected MSPEs.
 After $k_\vf+i$ Newton iterations we have $i$ bits of accuracy.
 The threshold $k_\vf$ depends on the representation size of $\vf$ and on the
 least solution $\mu\vf$. Although this latter dependence might seem to be a
 problem, lower and upper bounds on $\mu\vf$ can be
 easily derived for stochastic models (probabilistic
 programs with procedures, stochastic context-free grammars
 and back-button processes). In particular, this allows
 us to show that $k_\vf$ depends linearly on the representation
 size for back-button processes. We have also shown by means of an example
 that the threshold $k_\vf$ improves when
 the number of iterations increases.

 In \cite{KLE07:stoc} we
 left the problem whether DNM converges linearly for
 non-strongly-connected MSPEs open. We have proven that this is the case,
 although the convergence rate is poorer:
 if $h$ and $w$ are the height and width of the graph of SCCs of $\vf$,
 then there is a threshold
 $\widetilde k_\vf$ such that $\widetilde k_\vf + i \cdot w \cdot 2^{h+1}$
 iterations of DNM compute at least $i$ valid bits of
 $\fix{\vf}$,
 where the exponential factor cannot be avoided in general.

 Finally, we have shown that
 the Jacobian of the whole MSPE is guaranteed to exist,
 whether the MSPE is strongly connected or not.

 \section*{Acknowledgments.}
 \noindent The authors wish to thank Kousha Etessami and anonymous referees
 for very valuable comments. 

\bibliographystyle{plain} 

\end{document}